\documentclass[reprint,superscriptaddress,amsmath,amssymb,prl,floatfix]{revtex4-2}
\usepackage[dvipdfmx]{graphicx}
\usepackage{dcolumn}
\usepackage{bm}
\usepackage{xcolor}
\usepackage{multirow}
\usepackage{booktabs}

\usepackage{amsthm}
\theoremstyle{definition}

\usepackage{multibib}
\newcites{em}{End Matter References}

\begin{document}
\title{Statistical periodicity in noise-induced order from Ruelle--Pollicott resonances}

\author{Yuzuru Sato}
\email{ysato@math.sci.hokudai.ac.jp}
\affiliation{RIES-MSC, Hokkaido University, N12 W7 Kita-ku, Sapporo, 0600812 Hokkaido, Japan}
\affiliation{London Mathematical Laboratory, 14 Buckingham Street, London WC2N 6DF, United Kingdom}
\author{Isaia Nisoli}
\email{nisoli@im.ufrj.br}
\affiliation{Instituto de Matem\'atica, Universidade Federal do Rio de Janeiro, Av.\ Athos da Silveira Ramos 149, Rio de Janeiro, RJ 21941-909, Brazil}
\date{\today}

\begin{abstract}
Noise-induced order (NIO) is a paradigmatic example of nontrivial 
noise-induced phenomena, characterized by pronounced spectral peaks and
pseudoperiodic dynamics. We show that this periodicity is
governed by the Ruelle--Pollicott resonances of the annealed transfer
operator, independently of dynamical stability. Exact results for an
analytically solvable model and numerical results for a modified
Lasota--Mackey map show excellent agreement between resonance-based
predictions and empirical power spectra across a broad range of noise
amplitudes. Independent transitions in stability, diagnosed by the
Lyapunov exponent, and statistical periodicity, diagnosed by the
Ruelle--Pollicott resonances, give rise to three distinct types of NIO.

%We show that the Ruelle–Pollicott resonances of the transfer operator provide a universal mechanism for statistical periodicity in noise-induced order. 
%When subleading eigenvalues lie close to the unit circle, the power spectral density exhibits sharp peaks whose frequencies, heights, and linewidths are quantitatively determined by the spectral data, independently of the sign of the Lyapunov exponents. This mechanism accounts for the spectral sharpening observed in noise-induced order and extends beyond that setting, requiring only a spectral gap and complex eigenvalues. We further provide a time-series interpretation: eigenvalues govern the injection and expulsion rates of recurring, nearly periodic episodes, while eigenfunctions identify the corresponding oscillatory basins. These results are demonstrated analytically for an iterated function system and numerically for a modified Lasota–Mackey map, where the predicted power spectrum agrees with empirical spectra across multiple noise levels on both sides of the stability transitions.

\end{abstract}

\maketitle

% ======================================================================
% INTRODUCTION
% ======================================================================

Noise-induced phenomena emerge from the interplay between deterministic dynamics and stochastic perturbations in random dynamical systems~\cite{Arnold98}. Although such transitions arise across a broad range of applications, their theoretical foundations remain incomplete and many fundamental questions unresolved. A paradigmatic example is noise-induced order (NIO)~\cite{Matsumoto83}. In certain chaotic systems, increasing the noise amplitude drives the largest Lyapunov exponent from positive to negative values, reduces the Kolmogorov--Sinai entropy, generates pronounced peaks in the power spectrum, and induces pseudoperiodic dynamics. Attractor stability transitions and multiple transitions in NIO have been rigorously established using validated numerics~\cite{GalatoloEtAl2020,ChiharaEtAl2022}.
Yet the dynamical origin of this statistical pseudoperiodicity has remained elusive.

In this Letter, we refer to the pseudoperiodic dynamics
observed in NIO as \emph{statistical periodicity}, whereby individual
trajectories switch stochastically among multiple periodic patterns.
We show that this noise-induced behavior is governed by
Ruelle--Pollicott (RP) resonances~\cite{Ruelle86}, i.e., nontrivial isolated 
spectral values of the annealed transfer operator. We provide an explicit realization of Ruelle's spectral
picture~\cite{Ruelle86} for the pseudoperiodicity observed in
NIO~\cite{Matsumoto83}. 
In the present setting, the dominant nontrivial resonances form a
complex-conjugate pair, \(\lambda\) and \(\overline{\lambda}\). For the
discrete-time dynamics considered here, their argument determines the
characteristic period  
$T_{c}=2\pi/|\arg\lambda|$, whereas their modulus determines the average lifetime of the
corresponding periodic episodes  $\tau=|\lambda|/(1-|\lambda|)$. The spectral weights, and hence the
amplitudes of the associated peaks in the power spectrum, are determined
by the projections of the observable onto the corresponding eigenmodes. A defining feature of statistical periodicity is that pseudoperiodic
motion persists in the stationary regime, even though the ensemble
density converges to a stationary distribution. 
Statistical periodicity is therefore distinct from asymptotic
periodicity~\cite{Lasota87}, which is associated with nontrivial
peripheral eigenvalues on the unit circle, and from
metastability~\cite{MeynEtAl2008,MarcondesVaienti2026}, which is governed
by eigenvalues close to unity. It also differs from almost-cyclic
behavior~\cite{froyland2003detecting,SatoPadbergGehle2019}, which
describes oscillatory evolution of probability densities but does not,
by itself, characterize persistent pseudo-periodicity along individual
trajectories. Related spectral decompositions have also been developed
for stochastic differential equations~\cite{TantetEtAl2020a,
TantetEtAl2020b}. These distinctions are summarized in Table~I.

\begin{table}[htbp]
\small
\begin{tabular}{lccc}
\hline\hline
 & Asympt.\ per. & Metastab. & Stat.\ per. \\
\hline
Subleading eig. & $|\lambda| = 1$ & $\lambda \to 1$ & $|\lambda| < 1$ \\
Supports & disjoint & $\simeq$ disjoint & intermingled \\
Invariant\ density & cyclic & quasi-stat. & stationary \\
Power spectrum & $\delta$-peak & --- & Lorentzian \\
\hline\hline
\end{tabular}
\caption{
Comparison of asymptotic periodicity, metastability, and statistical
periodicity. $|\lambda|$ is the modulus of the dominant nontrivial eigenvalue.
%based on their subleading spectral values, support structure, invariant-density dynamics, and power spectra.
}
\label{tab:contrast}
\end{table}

%correlation decay
%measures the loss of memory of the membership of the oscillatory cluster 
%---how quickly an orbit forgets which oscillatory cluster it belongs to---
%not the dissolution of the periodic structure itself, much as
%marathon runners form persistent pace groups even though individual
%runners drift between them.

Notably, the complex RP resonances underlying this pseudoperiodicity
become dominant irrespective of the sign of the Lyapunov
exponent. We therefore classify NIO according to two independent
signatures: stabilization, determined by the Lyapunov exponent $\Lambda$, and statistical
periodicity, determined by a dominant complex-conjugate pair of RP
resonances $\lambda$ and $\overline{\lambda}$. The latter is characterized, for \(\sigma^2>0\), by the growth of a
pronounced spectral peak corresponding to a characteristic period
\(T_c\), which is inconspicuous in the deterministic limit, even though
the ensemble density converges to a stationary distribution.  Independent transitions
in stability and statistical periodicity  give rise to three distinct types of NIO. 
Type-I NIO exhibits stabilization without statistical periodicity;
type-II NIO exhibits statistical periodicity while remaining chaotic
($\Lambda>0$); and type-III NIO exhibits both stabilization and
statistical periodicity (see Table~II). Type-I NIO typically occurs in the strong-noise regime, where the random
dynamics can often be approximated by an effective linear Langevin
equation. By contrast, canonical examples of NIO, including the random
BZ map~\cite{Matsumoto83} and the random Lasota--Mackey
map~\cite{ChiharaEtAl2022}, belong to type III. The simple random maps
introduced below realize type-II NIO: statistical periodicity emerges, while the 
Lyapunov exponent remains positive.

\begin{table}[htbp]
\small
\begin{tabular}{lccc}
\hline\hline
 & Type-I NIO & Type-II NIO & Type-III NIO\\
\hline
Stability & $\Lambda_{\sigma}<0<\Lambda_{0}$ & $0<\Lambda_0, \Lambda_{\sigma} $ & $\Lambda_{\sigma}<0<\Lambda_{0}$\\
Char. per. & 
--- & $T_{c}$ \mbox{emerges}  & $T_{c}$ \mbox{emerges}\\
\hline\hline
\end{tabular}
\caption{
Comparison of type-I, type-II, and type-III NIO in terms of
noise-induced stabilization and noise-induced statistical periodicity with a characteristic
period \(T_c\). \(\Lambda_0\) and \(\Lambda_\sigma\) denote the
largest Lyapunov exponents in the deterministic and perturbed dynamical systems,
respectively.
}
\label{tab:contrast}
\end{table}

%We demonstrate the decoupling of spectral peaks from stability
%analytically---on an iterated function system with positive
%Lyapunov exponent and tunable periodicity---and numerically, by
%showing that the $\Lambda=0$ contour in parameter space is
%unrelated to the resonance structure.
%NIO is one manifestation of this general phenomenon.

A representative example of asymptotic periodicity is provided
by~\cite{Lasota87,Lasota91}
\[
x_{n+1}=\alpha x_n+d+\theta\eta_n \pmod 1,
\]
where \(\alpha\in[0,1)\), \(d\in\mathbb{R}\), and the \(\eta_n\) are
independent random variables uniformly distributed on \([0,1]\).
In the deterministic limit, \(\theta=0\), the system reduces to the
Nagumo--Sato map~\cite{nagumo1972response}. In both the deterministic
and stochastically perturbed cases, the transfer operator admits
asymptotically periodic densities. For \(\alpha=1/2\) and \(d=17/30\),
the deterministic map possesses an attracting period-3 orbit, while the
perturbed system exhibits an asymptotically period-3 density. Thus, the asymptotic periodicity is inherited from an attracting limit
cycle already present in the deterministic system
(\(\Lambda=\ln\alpha<0\)), rather than being generated by noise, and
therefore does not qualify as NIO.

%\textcolor{blue}{An example of meta-stability is given by ... (We give a simple example of meta-stability.)}

% ======================================================================
% ANALYTIC EXAMPLE: IFS
% ======================================================================

%{\bf Analytic example.}---
Here we introduce a minimal model of type-II NIO exhibiting statistical periodicity. Consider the random map
\begin{equation}
\begin{aligned}
x_{n+1}&=f(x_n), 
~\Pr(f=f_1)=p, ~\Pr(f=f_2)=1-p,\\
f_1(x)&=\frac{x-j/3}{1-3\epsilon}+\frac{j+1}{3}
\pmod 1, ~x\in I_j,\\
f_2(x)&=\frac{x-(j+1)/3}{1-3\epsilon}+\frac{j+2}{3}
\pmod 1, ~x\in I_j ,
\end{aligned}
\label{eq:rsm}
\end{equation}
where \(j=0,1,2\),
\(I_0=[0,1/3)\), \(I_1=[1/3,2/3)\), and \(I_2=[2/3,1]\),
with \(p\in[0,1]\) and \(0<\epsilon<1/6\). We refer to this model as the radio-station model.  The annealed Perron--Frobenius operator restricted to the invariant
subspace of densities that are piecewise constant on
\(\{I_0,I_1,I_2\}\) is represented by
\begin{equation}
P=
\begin{pmatrix}
3\epsilon(1-p) & 3\epsilon p & 1-3\epsilon\\
1-3\epsilon & 3\epsilon(1-p) & 3\epsilon p\\
3\epsilon p & 1-3\epsilon & 3\epsilon(1-p)
\end{pmatrix}.
\label{eq:rsm_matrix}
\end{equation}
Since \(P\) is circulant, its eigenvalues and corresponding right
eigenvectors are
\(\lambda_k=3\epsilon(1-p)+3\epsilon p\,\omega^k
+(1-3\epsilon)\omega^{2k}\) and
\(\bm v_k=(1,\omega^k,\omega^{2k})^T\), respectively, where
\(k=0,1,2\) and \(\omega=e^{2\pi i/3}\).
We have \(\lambda_0=1\) and
\(\lambda_2=\overline{\lambda_1}\), with
\(|\lambda_1|=|\lambda_2|<1\). The density therefore converges to the
uniform stationary state \(\bm\rho^*= \frac13\bm{1}\), with the decay rate
\(\kappa=-\ln|\lambda_1|\) and characteristic period
\(T_c=2\pi/|\arg\lambda_1|\). Let \(X_n\in\{0,1,2\}\) denote the interval index defined by
\(x_n\in I_{X_n}\), and consider the centered observable
\(\phi(X_n)=X_n-1\). Its stationary autocorrelation function is
\(C_\phi(n)=\frac23\operatorname{Re}(\lambda_1^n)\), corresponding to
exponentially damped oscillations. The discrete Wiener--Khinchin theorem
then gives the power spectrum
\begin{equation}
S(\theta)
=
\frac13
\left[
\frac{1-|\lambda_1|^2}
     {|1-\lambda_1e^{-i\theta}|^2}
+
\frac{1-|\lambda_1|^2}
     {|1-\overline{\lambda_1}e^{-i\theta}|^2}
\right],
\label{eq:psd_ifs}
\end{equation}
where $\theta\in[-\pi,\pi]$, which exhibits Lorentzian-like peaks near
\(\theta=\pm|\arg\lambda_1|\). Thus, the nontrivial RP resonances
determine the positions and widths of the spectral peaks, as well as the
pole structure of the power spectrum. By contrast, the maximal Lyapunov exponent is determined solely by the
common slope of the two maps and is given by
\(\Lambda=-\ln(1-3\epsilon)>0\), independently of \(p^*\). The dynamics
therefore remains chaotic, while the subleading RP resonances generate
statistical periodicity, providing an explicit realization of type-II
NIO. In the limit \(\epsilon\to0\), both maps reduce to a rigid period-3
rotation, and the power spectrum approaches
\(S(\theta)=(2\pi/3)
[\delta(\theta-2\pi/3)+\delta(\theta+2\pi/3)]\).
For the intermediate value \(\epsilon=1/12\), the characteristic period
is \(T_c\simeq3.568\) for \(p=0\) and \(T_c\simeq2.845\) for \(p=2/3\), whereas the Lyapunov exponent remains
\(\Lambda=\ln(4/3)\simeq0.2877>0\) in both cases.

%======================================================================
% MODEL: MODIFIED LASOTA-MACKEY MAP
% ======================================================================

%======================================================================
% SPECTRAL ANALYSIS
% ======================================================================
%
%{\bf Spectral analysis.}---
The annealed transfer operator of a random dynamical system with a map $f$ perturbed by additive Gaussian noise
acts on densities \(g\) as
\((\mathcal{L}_\sigma g)(x)
=\int g(y)\eta_\sigma(x-f(y))\,dy\),
where \(\eta_\sigma\) is the Gaussian kernel with variance
\(\sigma^2\). On a compact one-dimensional phase space, additive noise
with a bounded-variation density---including Gaussian and uniform
kernels, as well as piecewise-smooth kernels with finitely many
jumps---regularizes \(L^1\) densities into \(BV\). Consequently,
\(\mathcal{L}_\sigma\) is compact on \(L^1\), and its nonzero spectrum
consists of isolated eigenvalues of finite algebraic multiplicity, with
possible accumulation only at zero~\cite{Nisoli23,GalatoloEtAl2026}.
Let \(\lambda_0=1\) be a simple eigenvalue with associated strictly
positive stationary density \(\rho^*\), and let
\(\{\lambda_i\}_{i\geq1}\) denote the remaining nonzero eigenvalues,
ordered by decreasing modulus. Assuming these eigenvalues to be simple,
we denote the corresponding right and left eigenmodes by
\(\bm v_i\) and \(\bm u_i\), respectively, normalized such that
\(\langle\bm u_i,\bm v_j\rangle=\delta_{ij}\). On the zero-mean subspace, the spectral decomposition reads
\begin{equation}
\mathcal{L}_\sigma^n
=
\sum_{k=1}^{j}\lambda_k^n\Pi_k+R_j^n,
\qquad
\|R_j^n\|\leq Kr^n.
\end{equation}
where \(\Pi_k=\bm v_k\otimes\bm u_k\) are the spectral projectors, $K$ is a constant, and
\(R_j\) is the transient  operator. Here, \(r\) is chosen such that
\(r_{\mathrm{ess}}<r<|\lambda_j|<1\), where \(r_{\mathrm{ess}}\) is the essential spectral radius, and all eigenvalues satisfying
\(|\lambda_k| > r_{\mathrm{ess}}\) are included explicitly in the sum. The stationary
projector is \(\Pi_0=\bm v_0\otimes\bm u_0\), corresponding to the simple
eigenvalue \(\lambda_0=1\). 
For the observable \(\phi(x)=x\), define its centered version by
\(\bar\phi=\phi-\int\phi\,d\mu\), where
\(d\mu=\rho^*\,dm\). Its stationary autocorrelation function admits the
resonance expansion
\begin{equation}
C(n)
:=
\int \bar\phi\,
\mathcal{L}_\sigma^n(\bar\phi\,\rho)\,dm
=
\sum_{i\geq1}A_i(\phi)\lambda_i^n,
\label{eq:corr}
\end{equation}
where
\(A_i(\phi)
=\langle\bm u_i,\bar\phi\,\rho\rangle
 \int\bar\phi\,\bm v_i\,dm\)
is the spectral weight measuring the coupling of the observable to the
\(i\)th resonant mode. 
Each nonreal eigenvalue
\(\lambda_i=r_i e^{i\theta_i}\) contributes an exponentially
damped oscillatory mode with characteristic period
\(T_i=2\pi/|\theta_i|\). If \(r_i\) is interpreted as the
one-step survival probability of the corresponding periodic episode,
its average lifetime is
\(\tau_i=r_i/(1-r_i)\). 
The power spectrum can be written as 
\begin{equation}
S(\theta)\propto\frac{1}{2\pi}\sum_{k\ge1}
\left(
\frac{A_k(\phi)}{1-\lambda_k e^{-i\theta}}
+\frac{\overline{A_k(\phi)}}{1-\overline{\lambda_k} e^{-i\theta}}
\right)+E(\theta),
\label{eq:psd}
\end{equation}
where $E(\theta)$ is a bounded remainder from the essential spectrum (details in Appendix). A dominant eigenvalue $\lambda_1$ produces a peak at $\theta\simeq\pm\arg\lambda_1$ with height
\begin{equation}
S(|\arg\lambda_1|)\propto
\frac{2\,\mathrm{Re}(A_1)\,r_1}{1-r_1} = 2\mathrm{Re}(A_1)\tau_1.
\end{equation}
This relation explains the growth of
the dominant spectral peak observed in NIO~\cite{Matsumoto83}: its
resonant enhancement is controlled by the product of the episode
lifetime \(\tau_1\) and the spectral weight
\(2\,\mathrm{Re}[A_1(\phi)]\).

\begin{figure}[htbp]
    \centering
\includegraphics[width=\columnwidth]{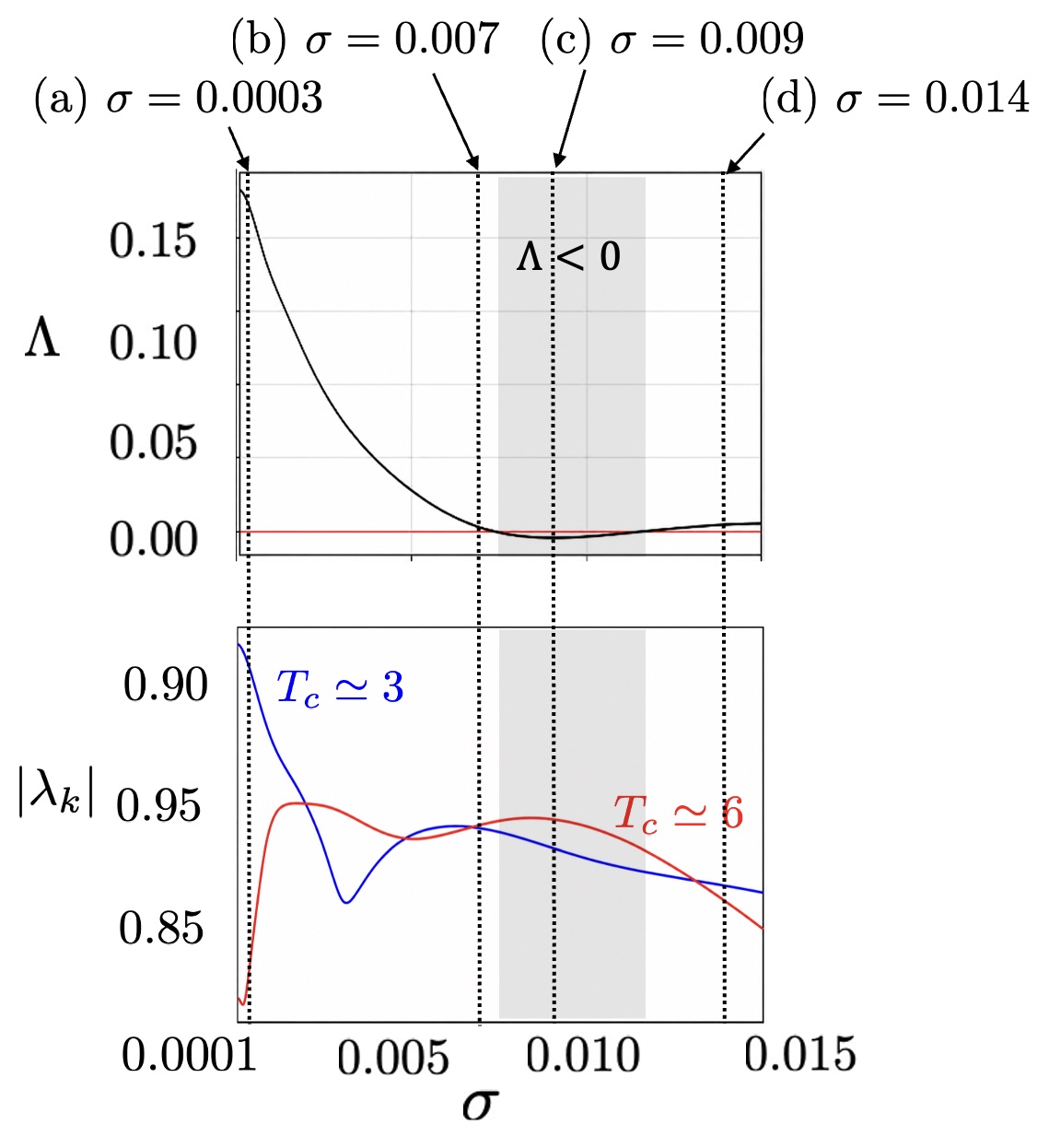}
    \caption{
    \textbf{Lyapunov exponents and Ruelle-Pollicott  resonances.} 
Lyapunov exponent \(\Lambda\) (top) and moduli of the dominant
RP resonances \(|\lambda_k|\) (bottom) as functions of the noise
amplitude \(\sigma\in[0.0001,0.015]\). The blue and red curves correspond
to resonant modes with characteristic periods \(T_c\simeq3\) and
\(T_c\simeq6\), respectively. The shaded region indicates
\(\Lambda<0\). The vertical dotted lines labeled (a)--(d) mark the noise
amplitudes used in Fig.~\ref{fig:statper}.
}
    \label{fig:lyapspec}
\end{figure}
\begin{figure}[htbp]
    \centering
\includegraphics[width=\columnwidth]{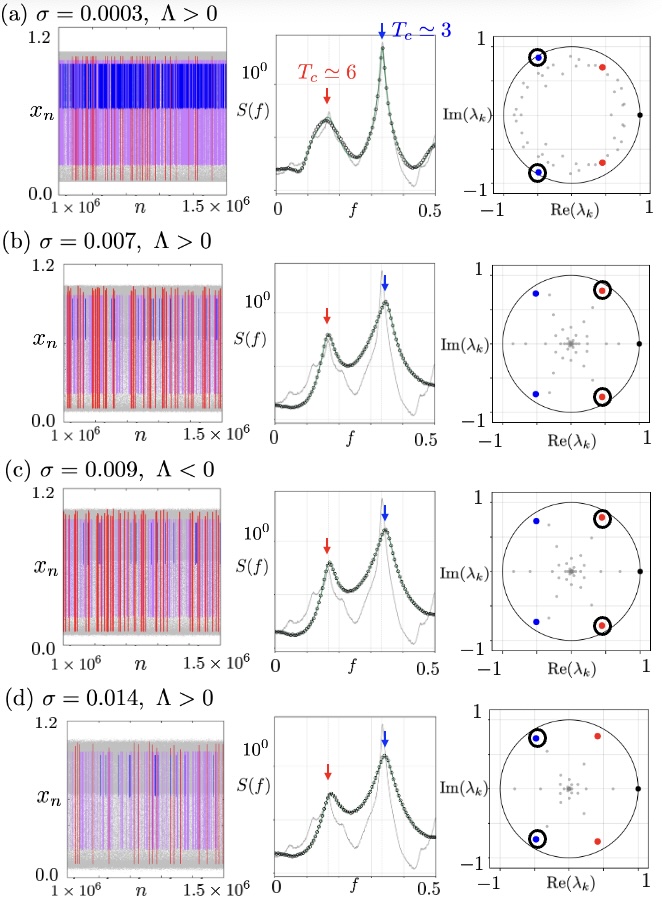}
    \caption{
\textbf{Statistical periodicity in the modified Lasota--Mackey map at
\(b=0.062\).}
Representative time series \(\{x_n\}\) (left), power spectra \(S(f)\)
(center), and spectra of the annealed transfer operator in the complex
plane (right) for
(a) \(\sigma=0.0003\), \(\Lambda>0\);
(b) \(\sigma=0.007\), \(\Lambda>0\);
(c) \(\sigma=0.009\), \(\Lambda<0\); and
(d) \(\sigma=0.014\), \(\Lambda>0\).
Trajectory segments associated with the period-3 UPOs are highlighted
in blue and purple, whereas those associated with the period-6 UPO are
highlighted in red; the remaining points are shown in gray (left).
The power spectra reconstructed from the RP-resonance expansion,
the empirical spectra computed from long trajectories, and the
deterministic spectra are shown as green curves, open circles, and gray
curves, respectively (center). Blue and red arrows mark the
characteristic frequencies corresponding to \(T_c\simeq3\) and
\(T_c\simeq6\), respectively. In the eigenvalue spectra, the blue and
red complex-conjugate pairs correspond to the period-3 and period-6
modes, respectively, while the remaining eigenvalues are shown as gray
dots (right). The dominant pair in each panel is enclosed by black
circles.
}
    \label{fig:statper}
\end{figure}

\begin{figure}[htbp]
    \centering
\includegraphics[width=0.98\columnwidth]{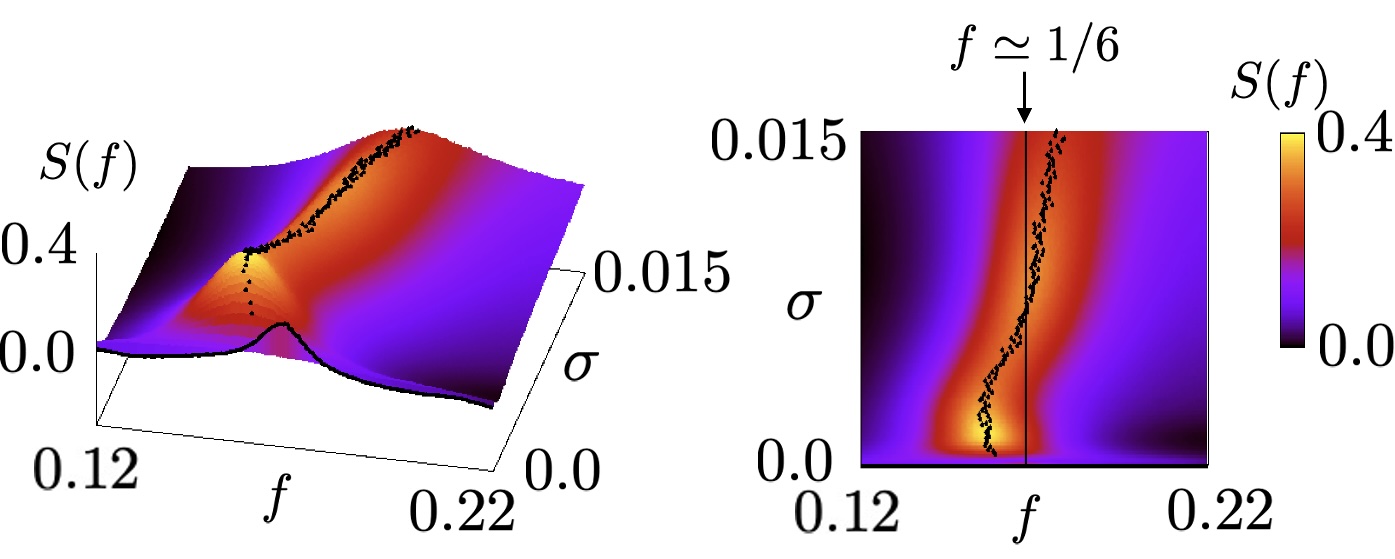}
\caption{
\textbf{Growth and shift of the period-6 spectral peak.}
The power spectrum \(S(f)\) is shown as a function of the noise
amplitude \(\sigma\), both as a three-dimensional surface (left) and as
a density plot (right). The black curve traces the spectral peak
associated with the period-6 mode. As \(\sigma\) increases, the peak
becomes more pronounced and shifts from \(T_c\simeq6.4\) in the
weak-noise regime to \(T_c\simeq5.7\) at \(\sigma=0.015\).
}
\label{fig:psd6}
\end{figure}

%{\bf Modified Lasota--Mackey map.}---
To investigate NIO in a more realistic setting, we consider the
stochastically perturbed map
\begin{eqnarray}
&&x_{n+1}=f(x_n)+\sigma\xi_n, \nonumber\\
&&f(x)=\alpha x+d-\frac{1}{1+e^{-\beta(\alpha x+d-1)}}+b,
\label{eq:mlm}
\end{eqnarray}
where \(\alpha=0.46\), \(d=17/30\), \(\beta=168\), \(b\) is a shift
parameter, and the \(\xi_n\sim\mathcal{N}(0,1)\) are i.i.d. Gaussian random variables; thus, \(\sigma\)
denotes the noise amplitude. This map is a sigmoidal smoothing of the
Nagumo--Sato neuron model~\cite{nagumo1972response,Aihara90}, whose
piecewise-linear form is recovered in the limit \(\beta\to\infty\).
For \(b=0.062\), the deterministic map exhibits chaotic dynamics,
whereas the addition of noise stabilizes the system,
driving the Lyapunov exponent \(\Lambda\) to negative
values~\cite{ChiharaEtAl2022,Nisoli23}. 
At the same time,  statistical
periodicity with characteristic periods near 3 and 6 occurs over the
range \(\sigma\in[0,0.015]\) (Fig.~\ref{fig:lyapspec}). As the noise
amplitude increases, the dominant RP resonances reorganize: the
complex-conjugate pairs associated with the period-3 and period-6 modes
exchange dominance, leading to a change in the dominant characteristic
period. In particular, the modulus of the period-3 pair initially
decreases and subsequently recovers. By contrast, the random Lyapunov
exponent \(\Lambda\) becomes negative at
\(\sigma\simeq0.0075\) and returns to positive values at
\(\sigma\simeq0.0115\). The mismatch between these stability transitions
and the resonance crossing demonstrates that statistical periodicity
and dynamical stability undergo distinct transitions.

Figure~\ref{fig:statper} shows representative time series, power
spectra, and spectra of the annealed transfer operator for \(b=0.062\)
and \(\sigma=0.0003\), \(0.007\), \(0.009\), and \(0.014\).
The trajectories frequently visit the neighborhoods of the period-3 UPO
\((0.62264,0.91508,0.93879)\) and the period-6 UPO
\((0.10482, 0.67688, 0.94003, 0.59955, 0.90446, 0.99272)\),
highlighted in blue and red, respectively, in the left panels. We also
identify another frequently visited period-3 UPO,
\((0.22099,0.73033,0.96462)\), highlighted in purple, whose
visitation frequency is largely insensitive to weak noise. Visits to
the neighborhoods of these UPOs generate the observed period-3 and
period-6 episodes. The power spectra reconstructed from the resonance
expansion in Eq.~(\ref{eq:psd}) (green curves) agree closely with the
empirical spectra computed from long simulated trajectories (open
circles), with peaks at 
\(\theta_i=|\arg\lambda_i|\) (middle panels). The deterministic power
spectrum is shown in gray for comparison. The right panels display the
eigenvalue spectra of the annealed transfer operator, revealing dominant
complex-conjugate pairs near the unit circle.

Figure~\ref{fig:psd6} illustrates the noise-induced growth and shift of
the spectral peak associated with the period-6 mode. As the noise
amplitude \(\sigma\) increases, the peak becomes more pronounced and
shifts from \(T_c\simeq6.4\) in the weak-noise regime to
\(T_c\simeq5.7\) at \(\sigma=0.015\).
Details of the numerical methods are provided in the
Appendix.

Figure~\ref{fig:phase} shows the phase diagram of the modified
Lasota--Mackey map in the \((b,\sigma)\) plane. Blue and red indicate
regions with dominant characteristic periods \(T_c\simeq3\) and
\(T_c\simeq6\), respectively. The random Lyapunov exponent \(\Lambda\)
changes sign along the black curve, whose geometry does not coincide
with the boundaries separating the dominant RP-resonance modes.
Pronounced spectral peaks occur on both sides of the
\(\Lambda=0\) contour, demonstrating that statistical periodicity is
independent of the Lyapunov-stability transition. Type-I, type-II, and
type-III NIO all occur within this parameter region, as indicated by the
white arrows.

%$|\lambda_2|$
%(panel~a), $|\arg(\lambda_2)|$ (panel~b), and the PSD peak height
%(panel~c) have their own, independent parameter dependence.

%\begin{figure}[tb]
%\centering
%\includegraphics[width=0.32\columnwidth]{absLambda2_LM.png}
%\includegraphics[width=0.32\columnwidth]{angle2_LM.png}
%\includegraphics[width=0.32\columnwidth]{peak2_LM.png}
%\caption{Parameter landscape for the modified Lasota--Mackey map
%($b\in[0.06,0.064]$, $\sigma\in[0.001,0.015]$).
%Black contour: $\Lambda=0$.
%(a)~$|\lambda_2|$;
%(b)~$|\arg(\lambda_2)|$;
%(c)~$\log_{10}$ PSD peak height $2|\mathrm{Re}(A_2)|\,|\lambda_2|/(1-|\lambda_2|)$.
%The resonance structure is independent of the Lyapunov zero contour.}
%\label{fig:landscape}
%\end{figure}

%\begin{figure}[tb]
%\centering
%\includegraphics[width=\columnwidth]{fig1.jpg}
%\caption{Statistical periodicity}
%\label{fig:statper}
%\end{figure}

%\begin{figure}[tb]
%\centering
%\includegraphics[width=\columnwidth]{fig2.jpg}
%\caption{Spectral analysis}
%\label{fig:specanal}
%\end{figure}

\begin{figure}[tb]
\centering
\includegraphics[width=\columnwidth]{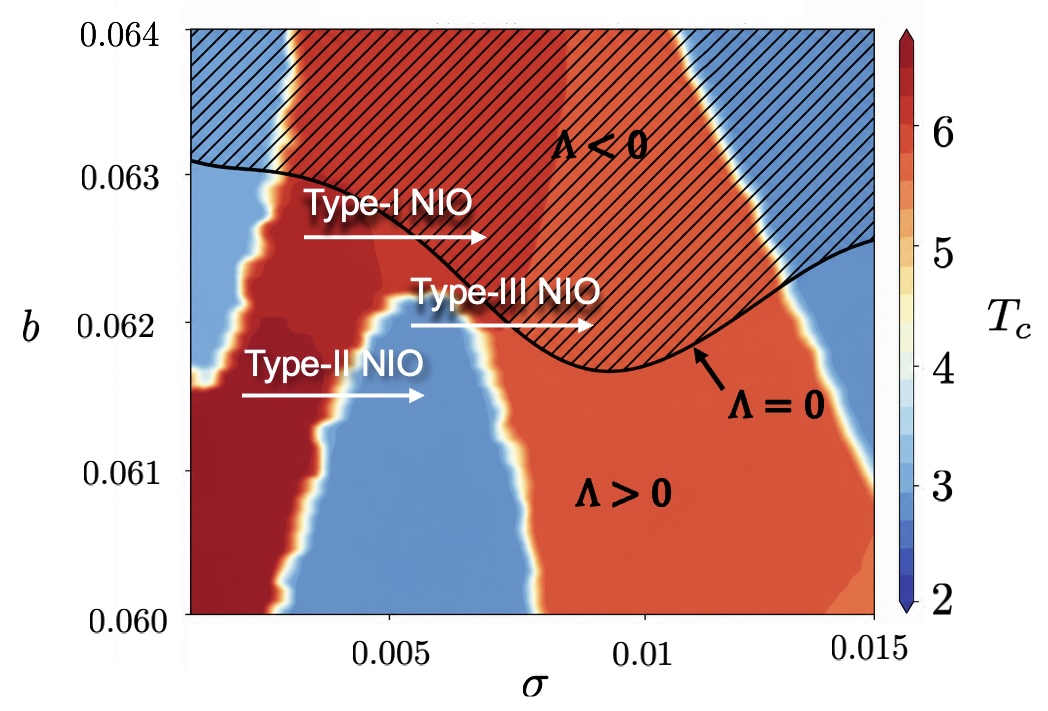}
\caption{
\textbf{Phase diagram of the modified Lasota--Mackey map in the
\((\sigma,b)\) plane.}
The color scale indicates the characteristic period
\(T_c=2\pi/|\arg\lambda|\), where \(\lambda\) denotes the dominant
complex RP resonance of the annealed transfer operator. The solid black
curve marks the Lyapunov-stability boundary \(\Lambda=0\); the hatched
and unhatched regions correspond to \(\Lambda<0\) and \(\Lambda>0\),
respectively. The white arrows identify representative parameter
regions exhibiting type-I, type-II, and type-III NIO.
}
\label{fig:phase}
\end{figure}

%\begin{figure}[tb]
%\centering
%\includegraphics[width=\columnwidth]{fig1.jpg}
%\caption{Spectral analysis at $b=0.062$, $\sigma=0.005$
%(Lyapunov exponent $\Lambda>0$).
%Top left: time series.
%Top right: PSD from time series (black) vs.\ RP resonance
%prediction (red); dashed lines mark $\omega=|\arg(\lambda_i)|$.
%Bottom left: eigenvalue spectrum on the unit disk.
%Bottom center: eigenfunction $\mathrm{Re}(f_2)$ (red) and
%invariant density $\rho$ (dashed).
%Right: slow-mode phase portrait
%$(\mathrm{Re}(f_2(x_n)),\mathrm{Im}(f_2(x_n)))$, showing the
%rotation-contraction by $\lambda_2$.
%See Supplemental Material Fig.~S11 for the same analysis at
%$\Lambda<0$.}
%\label{fig:psd}
%\end{figure}

% ======================================================================
% DISCUSSION
% ======================================================================

\vspace{2mm}

We have shown that statistical periodicity in random dynamical systems is governed by the Ruelle--Pollicott resonances of the annealed transfer
operator. The Lyapunov exponent
\(\Lambda=\int \log|f'|\,d\mu\) measures local dynamical instability by
averaging derivative information over the stationary measure, whereas
the subleading eigenvalues encode the decay and oscillation of
nonstationary modes. The radio-station model makes this distinction
explicit: \(\Lambda\) depends only on the common slope of the
piecewise-linear maps, whereas \(\arg\lambda_1\) is determined by the
transition structure among the partition elements. The phase diagram
of the modified Lasota--Mackey map further demonstrates that changes in
the dominant resonance structure do not follow the
\(\Lambda=0\) contour. Independent transitions in dynamical stability and statistical
periodicity, characterized by the Lyapunov exponent and the RP 
resonances, respectively, define three types of NIO: types I, II, and
III.
%This spectral perspective distinguishes noise-induced order from
%noise-induced stabilization and provides a general framework for
%analyzing emergent pseudoperiodicity in stochastic dynamics. 

Unlike asymptotic periodicity~\cite{Lasota87} and almost-cyclic
behavior~\cite{froyland2003detecting,SatoPadbergGehle2019}, the
statistical periodicity observed here does not rely on a cyclic
decomposition into disjoint or approximately disjoint supports.
Asymptotic periodicity is associated with densities that evolve
cyclically among disjoint supports, whereas almost-cyclic behavior
describes approximately cyclic transport among coherent sets. 
In the metastability literature~\cite{BovierDenHollander2015,
MarcondesVaienti2026}, the dominant subleading eigenvalues are typically
real and close to unity, resulting in long residence times. 
%but no finite-frequency oscillations or associated peaks in the power spectrum. 
These distinctions are summarized in Table~I. By contrast, the time series considered here exhibit recurrent episodes
of nearly periodic motion. Under a geometric-survival interpretation,
the dominant eigenvalues $\lambda_i$ gives the average 
lifetime
\(\tau_i=|\lambda_i|/(1-|\lambda_i|)\) of the corresponding periodic
episodes, while the spectral weight \(A_i(\phi)\), determined by the
projector \(\Pi_i\), quantifies the coupling of the observable to that
resonant mode. In our setting, the periodic episodes organized around
a small number of UPOs are intermingled, the stationary density is
unique, and the resonance expansion~(\ref{eq:psd}) provides quantitative
predictions for the power spectrum. 
%independently of dynamical stability. 
The Lorentzian spectral decompositions developed
in~\cite{TantetEtAl2020a,TantetEtAl2020b} are mathematically analogous
to Eq.~(\ref{eq:psd}). Here, we connect this spectral structure
explicitly to periodic episodes along individual
trajectories and show that it is independent of dynamical stability characterized by the Lyapunov exponent
\(\Lambda\). On a compact phase space, sufficiently regular additive
noise renders the transfer operator compact, with a purely discrete
nonzero spectrum, even for nonuniformly hyperbolic
dynamics~\cite{Nisoli23,GalatoloEtAl2026}. Because the mechanism depends
only on the spectral data of the annealed transfer operator, the present
theory and computational methods extend naturally to a broad class of chaotic 
random dynamical systems ~\cite{BlumenthalEtAl2026}.

\begin{acknowledgments}
Y.S.\ was supported by JSPS KAKENHI Grant No.\ JP21H01002 and JSPS Moonshot under Project No. JPMJMS2282-15. 
I.N.\ was partially supported by the Graduate Program in Mathematics
at UFRJ; CAPES, Finance Code 001; CNPq Universal Project
No.\ 404943/2023-3; CAPES--PRINT Grant No.\ 88881.311616/2018-00;
and CAPES--STINT Grant No.\ 88887.155746/2017-00.
\end{acknowledgments}

\bibliography{lmm}

%apsrev4-2.bst 2019-01-14 (MD) hand-edited version of apsrev4-1.bst
%Control: key (0)
%Control: author (8) initials jnrlst
%Control: editor formatted (1) identically to author
%Control: production of article title (0) allowed
%Control: page (0) single
%Control: year (1) truncated
%Control: production of eprint (0) enabled
\begin{thebibliography}{21}%
\makeatletter
\providecommand \@ifxundefined [1]{%
 \@ifx{#1\undefined}
}%
\providecommand \@ifnum [1]{%
 \ifnum #1\expandafter \@firstoftwo
 \else \expandafter \@secondoftwo
 \fi
}%
\providecommand \@ifx [1]{%
 \ifx #1\expandafter \@firstoftwo
 \else \expandafter \@secondoftwo
 \fi
}%
\providecommand \natexlab [1]{#1}%
\providecommand \enquote  [1]{``#1''}%
\providecommand \bibnamefont  [1]{#1}%
\providecommand \bibfnamefont [1]{#1}%
\providecommand \citenamefont [1]{#1}%
\providecommand \href@noop [0]{\@secondoftwo}%
\providecommand \href [0]{\begingroup \@sanitize@url \@href}%
\providecommand \@href[1]{\@@startlink{#1}\@@href}%
\providecommand \@@href[1]{\endgroup#1\@@endlink}%
\providecommand \@sanitize@url [0]{\catcode `\\12\catcode `\$12\catcode
  `\&12\catcode `\#12\catcode `\^12\catcode `\_12\catcode `\%12\relax}%
\providecommand \@@startlink[1]{}%
\providecommand \@@endlink[0]{}%
\providecommand \url  [0]{\begingroup\@sanitize@url \@url }%
\providecommand \@url [1]{\endgroup\@href {#1}{\urlprefix }}%
\providecommand \urlprefix  [0]{URL }%
\providecommand \Eprint [0]{\href }%
\providecommand \doibase [0]{https://doi.org/}%
\providecommand \selectlanguage [0]{\@gobble}%
\providecommand \bibinfo  [0]{\@secondoftwo}%
\providecommand \bibfield  [0]{\@secondoftwo}%
\providecommand \translation [1]{[#1]}%
\providecommand \BibitemOpen [0]{}%
\providecommand \bibitemStop [0]{}%
\providecommand \bibitemNoStop [0]{.\EOS\space}%
\providecommand \EOS [0]{\spacefactor3000\relax}%
\providecommand \BibitemShut  [1]{\csname bibitem#1\endcsname}%
\let\auto@bib@innerbib\@empty
%</preamble>
\bibitem [{\citenamefont {Arnold}(1998)}]{Arnold98}%
  \BibitemOpen
  \bibfield  {author} {\bibinfo {author} {\bibfnamefont {L.}~\bibnamefont
  {Arnold}},\ }\href@noop {} {\emph {\bibinfo {title} {Random Dynamical
  Systems}}}\ (\bibinfo  {publisher} {Springer-Verlag, Berlin},\ \bibinfo
  {year} {1998})\BibitemShut {NoStop}%
\bibitem [{\citenamefont {Matsumoto}\ and\ \citenamefont
  {Tsuda}(1983)}]{Matsumoto83}%
  \BibitemOpen
  \bibfield  {author} {\bibinfo {author} {\bibfnamefont {K.}~\bibnamefont
  {Matsumoto}}\ and\ \bibinfo {author} {\bibfnamefont {I.}~\bibnamefont
  {Tsuda}},\ }\bibfield  {title} {\bibinfo {title} {Noise induced order},\
  }\href@noop {} {\bibfield  {journal} {\bibinfo  {journal} {Journal of
  Statistical Physics}\ }\textbf {\bibinfo {volume} {31}},\ \bibinfo {pages}
  {87} (\bibinfo {year} {1983})}\BibitemShut {NoStop}%
\bibitem [{\citenamefont {Galatolo}\ \emph {et~al.}(2020)\citenamefont
  {Galatolo}, \citenamefont {Monge},\ and\ \citenamefont
  {Nisoli}}]{GalatoloEtAl2020}%
  \BibitemOpen
  \bibfield  {author} {\bibinfo {author} {\bibfnamefont {S.}~\bibnamefont
  {Galatolo}}, \bibinfo {author} {\bibfnamefont {M.}~\bibnamefont {Monge}},\
  and\ \bibinfo {author} {\bibfnamefont {I.}~\bibnamefont {Nisoli}},\
  }\bibfield  {title} {\bibinfo {title} {Existence of noise induced order, a
  computer aided proof},\ }\href@noop {} {\bibfield  {journal} {\bibinfo
  {journal} {Nonlinearity}\ }\textbf {\bibinfo {volume} {33}},\ \bibinfo
  {pages} {4237} (\bibinfo {year} {2020})}\BibitemShut {NoStop}%
\bibitem [{\citenamefont {Chihara}\ \emph {et~al.}(2022)\citenamefont
  {Chihara}, \citenamefont {Galatolo}, \citenamefont {Nisoli},\ and\
  \citenamefont {Sato}}]{ChiharaEtAl2022}%
  \BibitemOpen
  \bibfield  {author} {\bibinfo {author} {\bibfnamefont {T.}~\bibnamefont
  {Chihara}}, \bibinfo {author} {\bibfnamefont {S.}~\bibnamefont {Galatolo}},
  \bibinfo {author} {\bibfnamefont {I.}~\bibnamefont {Nisoli}},\ and\ \bibinfo
  {author} {\bibfnamefont {Y.}~\bibnamefont {Sato}},\ }\bibfield  {title}
  {\bibinfo {title} {Existence of multiple noise induced transitions in
  {L}asota--{M}ackey maps},\ }\href@noop {} {\bibfield  {journal} {\bibinfo
  {journal} {Nonlinearity}\ }\textbf {\bibinfo {volume} {35}},\ \bibinfo
  {pages} {5135} (\bibinfo {year} {2022})}\BibitemShut {NoStop}%
\bibitem [{\citenamefont {Ruelle}(1986)}]{Ruelle86}%
  \BibitemOpen
  \bibfield  {author} {\bibinfo {author} {\bibfnamefont {D.}~\bibnamefont
  {Ruelle}},\ }\bibfield  {title} {\bibinfo {title} {Resonances of chaotic
  dynamical systems},\ }\href@noop {} {\bibfield  {journal} {\bibinfo
  {journal} {Physical Review Letters}\ }\textbf {\bibinfo {volume} {56}},\
  \bibinfo {pages} {405} (\bibinfo {year} {1986})}\BibitemShut {NoStop}%
\bibitem [{\citenamefont {Lasota}\ and\ \citenamefont
  {Mackey}(1987)}]{Lasota87}%
  \BibitemOpen
  \bibfield  {author} {\bibinfo {author} {\bibfnamefont {A.}~\bibnamefont
  {Lasota}}\ and\ \bibinfo {author} {\bibfnamefont {M.}~\bibnamefont
  {Mackey}},\ }\bibfield  {title} {\bibinfo {title} {Noise and statistical
  periodicity},\ }\href@noop {} {\bibfield  {journal} {\bibinfo  {journal}
  {Physica}\ }\textbf {\bibinfo {volume} {D 28}},\ \bibinfo {pages} {143}
  (\bibinfo {year} {1987})}\BibitemShut {NoStop}%
\bibitem [{\citenamefont {Meyn}\ \emph {et~al.}(2008)\citenamefont {Meyn},
  \citenamefont {Hagen}, \citenamefont {Mathew},\ and\ \citenamefont
  {Banaszuk}}]{MeynEtAl2008}%
  \BibitemOpen
  \bibfield  {author} {\bibinfo {author} {\bibfnamefont {S.}~\bibnamefont
  {Meyn}}, \bibinfo {author} {\bibfnamefont {G.}~\bibnamefont {Hagen}},
  \bibinfo {author} {\bibfnamefont {G.}~\bibnamefont {Mathew}},\ and\ \bibinfo
  {author} {\bibfnamefont {A.}~\bibnamefont {Banaszuk}},\ }\bibfield  {title}
  {\bibinfo {title} {On complex spectra and metastability of {M}arkov models},\
  }in\ \href@noop {} {\emph {\bibinfo {booktitle} {Proceedings of the 47th IEEE
  Conference on Decision and Control}}}\ (\bibinfo  {publisher} {IEEE},\
  \bibinfo {address} {Canc\'un},\ \bibinfo {year} {2008})\ pp.\ \bibinfo
  {pages} {2192--2197}\BibitemShut {NoStop}%
\bibitem [{\citenamefont {Marcondes}\ and\ \citenamefont
  {Vaienti}(2026)}]{MarcondesVaienti2026}%
  \BibitemOpen
  \bibfield  {author} {\bibinfo {author} {\bibfnamefont {D.}~\bibnamefont
  {Marcondes}}\ and\ \bibinfo {author} {\bibfnamefont {S.}~\bibnamefont
  {Vaienti}},\ }\href@noop {} {\bibinfo {title} {Metastability of random maps:
  a resolvent approach}} (\bibinfo {year} {2026}),\ \Eprint
  {https://arxiv.org/abs/2602.12400} {arXiv:2602.12400 [math.DS]} \BibitemShut
  {NoStop}%
\bibitem [{\citenamefont {Froyland}\ and\ \citenamefont
  {Dellnitz}(2003)}]{froyland2003detecting}%
  \BibitemOpen
  \bibfield  {author} {\bibinfo {author} {\bibfnamefont {G.}~\bibnamefont
  {Froyland}}\ and\ \bibinfo {author} {\bibfnamefont {M.}~\bibnamefont
  {Dellnitz}},\ }\bibfield  {title} {\bibinfo {title} {Detecting and locating
  near-optimal almost-invariant sets and cycles},\ }\href@noop {} {\bibfield
  {journal} {\bibinfo  {journal} {SIAM Journal on Scientific Computing}\
  }\textbf {\bibinfo {volume} {24}},\ \bibinfo {pages} {1839} (\bibinfo {year}
  {2003})}\BibitemShut {NoStop}%
\bibitem [{\citenamefont {Sato}\ and\ \citenamefont
  {Padberg-Gehle}(2019)}]{SatoPadbergGehle2019}%
  \BibitemOpen
  \bibfield  {author} {\bibinfo {author} {\bibfnamefont {Y.}~\bibnamefont
  {Sato}}\ and\ \bibinfo {author} {\bibfnamefont {K.}~\bibnamefont
  {Padberg-Gehle}},\ }\href@noop {} {\bibinfo {title} {Noise-induced
  statistical periodicity in random {L}asota--{M}ackey maps}} (\bibinfo {year}
  {2019}),\ \Eprint {https://arxiv.org/abs/1905.02746} {arXiv:1905.02746
  [nlin.CD]} \BibitemShut {NoStop}%
\bibitem [{\citenamefont {Tantet}\ \emph
  {et~al.}(2020{\natexlab{a}})\citenamefont {Tantet}, \citenamefont {Chekroun},
  \citenamefont {Dijkstra},\ and\ \citenamefont {Neelin}}]{TantetEtAl2020a}%
  \BibitemOpen
  \bibfield  {author} {\bibinfo {author} {\bibfnamefont {A.}~\bibnamefont
  {Tantet}}, \bibinfo {author} {\bibfnamefont {M.~D.}\ \bibnamefont
  {Chekroun}}, \bibinfo {author} {\bibfnamefont {H.~A.}\ \bibnamefont
  {Dijkstra}},\ and\ \bibinfo {author} {\bibfnamefont {J.~D.}\ \bibnamefont
  {Neelin}},\ }\bibfield  {title} {\bibinfo {title} {Ruelle--{P}ollicott
  resonances of stochastic systems in reduced state space. {P}art {I}:
  {T}heory},\ }\href@noop {} {\bibfield  {journal} {\bibinfo  {journal}
  {Journal of Statistical Physics}\ }\textbf {\bibinfo {volume} {179}},\
  \bibinfo {pages} {1366} (\bibinfo {year} {2020}{\natexlab{a}})}\BibitemShut
  {NoStop}%
\bibitem [{\citenamefont {Tantet}\ \emph
  {et~al.}(2020{\natexlab{b}})\citenamefont {Tantet}, \citenamefont {Chekroun},
  \citenamefont {Dijkstra},\ and\ \citenamefont {Neelin}}]{TantetEtAl2020b}%
  \BibitemOpen
  \bibfield  {author} {\bibinfo {author} {\bibfnamefont {A.}~\bibnamefont
  {Tantet}}, \bibinfo {author} {\bibfnamefont {M.~D.}\ \bibnamefont
  {Chekroun}}, \bibinfo {author} {\bibfnamefont {H.~A.}\ \bibnamefont
  {Dijkstra}},\ and\ \bibinfo {author} {\bibfnamefont {J.~D.}\ \bibnamefont
  {Neelin}},\ }\bibfield  {title} {\bibinfo {title} {Ruelle--{P}ollicott
  resonances of stochastic systems in reduced state space. {P}art {II}:
  {S}tochastic {H}opf bifurcation},\ }\href@noop {} {\bibfield  {journal}
  {\bibinfo  {journal} {Journal of Statistical Physics}\ }\textbf {\bibinfo
  {volume} {179}},\ \bibinfo {pages} {1403} (\bibinfo {year}
  {2020}{\natexlab{b}})}\BibitemShut {NoStop}%
\bibitem [{\citenamefont {Lasota}\ and\ \citenamefont
  {Mackey}(1991)}]{Lasota91}%
  \BibitemOpen
  \bibfield  {author} {\bibinfo {author} {\bibfnamefont {A.}~\bibnamefont
  {Lasota}}\ and\ \bibinfo {author} {\bibfnamefont {M.}~\bibnamefont
  {Mackey}},\ }\href@noop {} {\emph {\bibinfo {title} {Chaos, Fractals, and
  Noise}}}\ (\bibinfo  {publisher} {Springer},\ \bibinfo {year}
  {1991})\BibitemShut {NoStop}%
\bibitem [{\citenamefont {Nagumo}\ and\ \citenamefont
  {Sato}(1972)}]{nagumo1972response}%
  \BibitemOpen
  \bibfield  {author} {\bibinfo {author} {\bibfnamefont {J.}~\bibnamefont
  {Nagumo}}\ and\ \bibinfo {author} {\bibfnamefont {S.}~\bibnamefont {Sato}},\
  }\bibfield  {title} {\bibinfo {title} {On a response characteristic of a
  mathematical neuron model},\ }\href@noop {} {\bibfield  {journal} {\bibinfo
  {journal} {Kybernetik}\ }\textbf {\bibinfo {volume} {10}},\ \bibinfo {pages}
  {155} (\bibinfo {year} {1972})}\BibitemShut {NoStop}%
\bibitem [{\citenamefont {Nisoli}(2023)}]{Nisoli23}%
  \BibitemOpen
  \bibfield  {author} {\bibinfo {author} {\bibfnamefont {I.}~\bibnamefont
  {Nisoli}},\ }\bibfield  {title} {\bibinfo {title} {How does noise induce
  order?},\ }\href@noop {} {\bibfield  {journal} {\bibinfo  {journal} {Journal
  of Statistical Physics}\ }\textbf {\bibinfo {volume} {190}},\ \bibinfo
  {pages} {22} (\bibinfo {year} {2023})}\BibitemShut {NoStop}%
\bibitem [{\citenamefont {Galatolo}\ \emph {et~al.}(2026)\citenamefont
  {Galatolo}, \citenamefont {Vereau}, \citenamefont {Marangio},\ and\
  \citenamefont {Nisoli}}]{GalatoloEtAl2026}%
  \BibitemOpen
  \bibfield  {author} {\bibinfo {author} {\bibfnamefont {S.}~\bibnamefont
  {Galatolo}}, \bibinfo {author} {\bibfnamefont {C.~L.}\ \bibnamefont
  {Vereau}}, \bibinfo {author} {\bibfnamefont {L.}~\bibnamefont {Marangio}},\
  and\ \bibinfo {author} {\bibfnamefont {I.}~\bibnamefont {Nisoli}},\
  }\href@noop {} {\bibinfo {title} {Efficient computation of stationary
  measures and the {L}yapunov landscape for families of random dynamical
  systems with smooth additive noise}} (\bibinfo {year} {2026}),\ \Eprint
  {https://arxiv.org/abs/2508.03895} {arXiv:2508.03895 [math.DS]} \BibitemShut
  {NoStop}%
\bibitem [{\citenamefont {Aihara}\ \emph {et~al.}(1990)\citenamefont {Aihara},
  \citenamefont {Takabe},\ and\ \citenamefont {Toyoda}}]{Aihara90}%
  \BibitemOpen
  \bibfield  {author} {\bibinfo {author} {\bibfnamefont {K.}~\bibnamefont
  {Aihara}}, \bibinfo {author} {\bibfnamefont {T.}~\bibnamefont {Takabe}},\
  and\ \bibinfo {author} {\bibfnamefont {M.}~\bibnamefont {Toyoda}},\
  }\bibfield  {title} {\bibinfo {title} {Chaotic neural networks},\ }\href@noop
  {} {\bibfield  {journal} {\bibinfo  {journal} {Physics Letters}\ }\textbf
  {\bibinfo {volume} {A 144}},\ \bibinfo {pages} {333} (\bibinfo {year}
  {1990})}\BibitemShut {NoStop}%
\bibitem [{\citenamefont {Bovier}\ and\ \citenamefont {den
  Hollander}(2015)}]{BovierDenHollander2015}%
  \BibitemOpen
  \bibfield  {author} {\bibinfo {author} {\bibfnamefont {A.}~\bibnamefont
  {Bovier}}\ and\ \bibinfo {author} {\bibfnamefont {F.}~\bibnamefont {den
  Hollander}},\ }\href@noop {} {\emph {\bibinfo {title} {Metastability: A
  Potential-Theoretic Approach}}},\ \bibinfo {series} {Grundlehren der
  mathematischen Wissenschaften}, Vol.\ \bibinfo {volume} {351}\ (\bibinfo
  {publisher} {Springer},\ \bibinfo {year} {2015})\BibitemShut {NoStop}%
\bibitem [{\citenamefont {Blumenthal}\ \emph {et~al.}(2026)\citenamefont
  {Blumenthal}, \citenamefont {Nisoli},\ and\ \citenamefont
  {Taylor-Crush}}]{BlumenthalEtAl2026}%
  \BibitemOpen
  \bibfield  {author} {\bibinfo {author} {\bibfnamefont {A.}~\bibnamefont
  {Blumenthal}}, \bibinfo {author} {\bibfnamefont {I.}~\bibnamefont {Nisoli}},\
  and\ \bibinfo {author} {\bibfnamefont {T.}~\bibnamefont {Taylor-Crush}},\
  }\href@noop {} {\bibinfo {title} {A pseudospectral approach to rigorous
  numerical estimation of resonances of transfer operators}} (\bibinfo {year}
  {2026}),\ \bibinfo {note} {submitted}\BibitemShut {NoStop}%
\bibitem [{\citenamefont {Golub}\ and\ \citenamefont
  {Van~Loan}(2013)}]{GolubLoan2013}%
  \BibitemOpen
  \bibfield  {author} {\bibinfo {author} {\bibfnamefont {G.~H.}\ \bibnamefont
  {Golub}}\ and\ \bibinfo {author} {\bibfnamefont {C.~F.}\ \bibnamefont
  {Van~Loan}},\ }\href@noop {} {\emph {\bibinfo {title} {Matrix
  Computations}}},\ \bibinfo {edition} {4th}\ ed.\ (\bibinfo  {publisher}
  {Johns Hopkins University Press},\ \bibinfo {address} {Baltimore, MD},\
  \bibinfo {year} {2013})\BibitemShut {NoStop}%
\bibitem [{\citenamefont {Welch}(1967)}]{Welch1967}%
  \BibitemOpen
  \bibfield  {author} {\bibinfo {author} {\bibfnamefont {P.~D.}\ \bibnamefont
  {Welch}},\ }\bibfield  {title} {\bibinfo {title} {The use of fast {Fourier}
  transform for the estimation of power spectra: A method based on time
  averaging over short, modified periodograms},\ }\href
  {https://doi.org/10.1109/TAU.1967.1161901} {\bibfield  {journal} {\bibinfo
  {journal} {IEEE Transactions on Audio and Electroacoustics}\ }\textbf
  {\bibinfo {volume} {15}},\ \bibinfo {pages} {70} (\bibinfo {year}
  {1967})}\BibitemShut {NoStop}%
\end{thebibliography}%


%apsrev4-2.bst 2019-01-14 (MD) hand-edited version of apsrev4-1.bst
%Control: key (0)
%Control: author (72) initials jnrlst
%Control: editor formatted (1) identically to author
%Control: production of article title (-1) disabled
%Control: page (0) single
%Control: year (1) truncated
%Control: production of eprint (0) enabled
\begin{thebibliography}{12}%
\makeatletter
\providecommand \@ifxundefined [1]{%
 \@ifx{#1\undefined}
}%
\providecommand \@ifnum [1]{%
 \ifnum #1\expandafter \@firstoftwo
 \else \expandafter \@secondoftwo
 \fi
}%
\providecommand \@ifx [1]{%
 \ifx #1\expandafter \@firstoftwo
 \else \expandafter \@secondoftwo
 \fi
}%
\providecommand \natexlab [1]{#1}%
\providecommand \enquote  [1]{``#1''}%
\providecommand \bibnamefont  [1]{#1}%
\providecommand \bibfnamefont [1]{#1}%
\providecommand \citenamefont [1]{#1}%
\providecommand \href@noop [0]{\@secondoftwo}%
\providecommand \href [0]{\begingroup \@sanitize@url \@href}%
\providecommand \@href[1]{\@@startlink{#1}\@@href}%
\providecommand \@@href[1]{\endgroup#1\@@endlink}%
\providecommand \@sanitize@url [0]{\catcode `\\12\catcode `\$12\catcode
  `\&12\catcode `\#12\catcode `\^12\catcode `\_12\catcode `\%12\relax}%
\providecommand \@@startlink[1]{}%
\providecommand \@@endlink[0]{}%
\providecommand \url  [0]{\begingroup\@sanitize@url \@url }%
\providecommand \@url [1]{\endgroup\@href {#1}{\urlprefix }}%
\providecommand \urlprefix  [0]{URL }%
\providecommand \Eprint [0]{\href }%
\providecommand \doibase [0]{https://doi.org/}%
\providecommand \selectlanguage [0]{\@gobble}%
\providecommand \bibinfo  [0]{\@secondoftwo}%
\providecommand \bibfield  [0]{\@secondoftwo}%
\providecommand \translation [1]{[#1]}%
\providecommand \BibitemOpen [0]{}%
\providecommand \bibitemStop [0]{}%
\providecommand \bibitemNoStop [0]{.\EOS\space}%
\providecommand \EOS [0]{\spacefactor3000\relax}%
\providecommand \BibitemShut  [1]{\csname bibitem#1\endcsname}%
\let\auto@bib@innerbib\@empty
%</preamble>
\bibitem [{\citenamefont {Galatolo}\ \emph {et~al.}(2026)\citenamefont
  {Galatolo}, \citenamefont {Vereau}, \citenamefont {Marangio},\ and\
  \citenamefont {Nisoli}}]{GalatoloEtAl2026}%
  \BibitemOpen
  \bibfield  {author} {\bibinfo {author} {\bibfnamefont {S.}~\bibnamefont
  {Galatolo}}, \bibinfo {author} {\bibfnamefont {C.~L.}\ \bibnamefont
  {Vereau}}, \bibinfo {author} {\bibfnamefont {L.}~\bibnamefont {Marangio}},\
  and\ \bibinfo {author} {\bibfnamefont {I.}~\bibnamefont {Nisoli}},\
  }\href@noop {} {\bibinfo {title} {Efficient computation of stationary
  measures and the {L}yapunov landscape for families of random dynamical
  systems with smooth additive noise}} (\bibinfo {year} {2026}),\ \Eprint
  {https://arxiv.org/abs/2508.03895} {arXiv:2508.03895 [math.DS]} \BibitemShut
  {NoStop}%
\bibitem [{\citenamefont {Galatolo}\ \emph {et~al.}(2020)\citenamefont
  {Galatolo}, \citenamefont {Monge},\ and\ \citenamefont
  {Nisoli}}]{GalatoloEtAl2020}%
  \BibitemOpen
  \bibfield  {author} {\bibinfo {author} {\bibfnamefont {S.}~\bibnamefont
  {Galatolo}}, \bibinfo {author} {\bibfnamefont {M.}~\bibnamefont {Monge}},\
  and\ \bibinfo {author} {\bibfnamefont {I.}~\bibnamefont {Nisoli}},\
  }\href@noop {} {\bibfield  {journal} {\bibinfo  {journal} {Nonlinearity}\
  }\textbf {\bibinfo {volume} {33}},\ \bibinfo {pages} {4237} (\bibinfo {year}
  {2020})}\BibitemShut {NoStop}%
\bibitem [{\citenamefont {Nisoli}\ and\ \citenamefont
  {Poloni}(2024)}]{NisoliPoloni_RIM}%
  \BibitemOpen
  \bibfield  {author} {\bibinfo {author} {\bibfnamefont {I.}~\bibnamefont
  {Nisoli}}\ and\ \bibinfo {author} {\bibfnamefont {F.}~\bibnamefont
  {Poloni}},\ }\href
  {https://github.com/JuliaDynamics/RigorousInvariantMeasures.jl} {\bibinfo
  {title} {{RigorousInvariantMeasures.jl}}} (\bibinfo {year}
  {2024})\BibitemShut {NoStop}%
\bibitem [{\citenamefont {Ruelle}(1986)}]{Ruelle86}%
  \BibitemOpen
  \bibfield  {author} {\bibinfo {author} {\bibfnamefont {D.}~\bibnamefont
  {Ruelle}},\ }\href@noop {} {\bibfield  {journal} {\bibinfo  {journal}
  {Physical Review Letters}\ }\textbf {\bibinfo {volume} {56}},\ \bibinfo
  {pages} {405} (\bibinfo {year} {1986})}\BibitemShut {NoStop}%
\bibitem [{\citenamefont {Dyatlov}\ and\ \citenamefont
  {Zworski}(2015)}]{DyatlovZworski2015}%
  \BibitemOpen
  \bibfield  {author} {\bibinfo {author} {\bibfnamefont {S.}~\bibnamefont
  {Dyatlov}}\ and\ \bibinfo {author} {\bibfnamefont {M.}~\bibnamefont
  {Zworski}},\ }\href {https://doi.org/10.1088/0951-7715/28/10/3511} {\bibfield
   {journal} {\bibinfo  {journal} {Nonlinearity}\ }\textbf {\bibinfo {volume}
  {28}},\ \bibinfo {pages} {3511} (\bibinfo {year} {2015})}\BibitemShut
  {NoStop}%
\bibitem [{\citenamefont {Blumenthal}\ \emph {et~al.}(2026)\citenamefont
  {Blumenthal}, \citenamefont {Nisoli},\ and\ \citenamefont
  {Taylor-Crush}}]{BlumenthalEtAl2026}%
  \BibitemOpen
  \bibfield  {author} {\bibinfo {author} {\bibfnamefont {A.}~\bibnamefont
  {Blumenthal}}, \bibinfo {author} {\bibfnamefont {I.}~\bibnamefont {Nisoli}},\
  and\ \bibinfo {author} {\bibfnamefont {T.}~\bibnamefont {Taylor-Crush}},\
  }\href@noop {} {\bibinfo {title} {A pseudospectral approach to rigorous
  numerical estimation of resonances of transfer operators}} (\bibinfo {year}
  {2026}),\ \bibinfo {note} {submitted}\BibitemShut {NoStop}%
\bibitem [{\citenamefont {Bezanson}\ \emph {et~al.}(2017)\citenamefont
  {Bezanson}, \citenamefont {Edelman}, \citenamefont {Karpinski},\ and\
  \citenamefont {Shah}}]{Bezanson2017Julia}%
  \BibitemOpen
  \bibfield  {author} {\bibinfo {author} {\bibfnamefont {J.}~\bibnamefont
  {Bezanson}}, \bibinfo {author} {\bibfnamefont {A.}~\bibnamefont {Edelman}},
  \bibinfo {author} {\bibfnamefont {S.}~\bibnamefont {Karpinski}},\ and\
  \bibinfo {author} {\bibfnamefont {V.~B.}\ \bibnamefont {Shah}},\ }\href
  {https://doi.org/10.1137/141000671} {\bibfield  {journal} {\bibinfo
  {journal} {SIAM Review}\ }\textbf {\bibinfo {volume} {59}},\ \bibinfo {pages}
  {65} (\bibinfo {year} {2017})}\BibitemShut {NoStop}%
\bibitem [{\citenamefont {Sanders}\ and\ \citenamefont
  {Benet}(2014)}]{SandersBenet2014}%
  \BibitemOpen
  \bibfield  {author} {\bibinfo {author} {\bibfnamefont {D.~P.}\ \bibnamefont
  {Sanders}}\ and\ \bibinfo {author} {\bibfnamefont {L.}~\bibnamefont
  {Benet}},\ }\href {https://doi.org/10.5281/zenodo.3336308} {\bibinfo {title}
  {{IntervalArithmetic.jl}}} (\bibinfo {year} {2014})\BibitemShut {NoStop}%
\bibitem [{\citenamefont {Ferranti}\ and\ \citenamefont
  {Nisoli}(2024)}]{FerrantiNisoli_BallArithmetic}%
  \BibitemOpen
  \bibfield  {author} {\bibinfo {author} {\bibfnamefont {L.}~\bibnamefont
  {Ferranti}}\ and\ \bibinfo {author} {\bibfnamefont {I.}~\bibnamefont
  {Nisoli}},\ }\href {https://github.com/JuliaBallArithmetic/BallArithmetic.jl}
  {\bibinfo {title} {{BallArithmetic.jl}}} (\bibinfo {year} {2024})\BibitemShut
  {NoStop}%
\bibitem [{\citenamefont {Johnson}\ and\ \citenamefont
  {Frigo}(2008)}]{JohnsonFrigo2008}%
  \BibitemOpen
  \bibfield  {author} {\bibinfo {author} {\bibfnamefont {S.~G.}\ \bibnamefont
  {Johnson}}\ and\ \bibinfo {author} {\bibfnamefont {M.}~\bibnamefont
  {Frigo}},\ }in\ \href {http://cnx.org/content/m16336/1.14/} {\emph {\bibinfo
  {booktitle} {Fast Fourier Transforms}}},\ \bibinfo {editor} {edited by\
  \bibinfo {editor} {\bibfnamefont {C.~S.}\ \bibnamefont {Burrus}}}\ (\bibinfo
  {publisher} {Connexions, Rice University},\ \bibinfo {address} {Houston,
  TX},\ \bibinfo {year} {2008})\ Chap.~\bibinfo {chapter} {11}\BibitemShut
  {NoStop}%
\bibitem [{\citenamefont {Revels}\ \emph {et~al.}(2016)\citenamefont {Revels},
  \citenamefont {Lubin},\ and\ \citenamefont
  {Papamarkou}}]{Revels2016ForwardDiff}%
  \BibitemOpen
  \bibfield  {author} {\bibinfo {author} {\bibfnamefont {J.}~\bibnamefont
  {Revels}}, \bibinfo {author} {\bibfnamefont {M.}~\bibnamefont {Lubin}},\ and\
  \bibinfo {author} {\bibfnamefont {T.}~\bibnamefont {Papamarkou}},\ }\href
  {https://arxiv.org/abs/1607.07892} {\bibfield  {journal} {\bibinfo  {journal}
  {arXiv:1607.07892 [cs.MS]}\ } (\bibinfo {year} {2016})}\BibitemShut {NoStop}%
\bibitem [{\citenamefont {Danisch}\ and\ \citenamefont
  {Krumbiegel}(2021)}]{Danisch2021Makie}%
  \BibitemOpen
  \bibfield  {author} {\bibinfo {author} {\bibfnamefont {S.}~\bibnamefont
  {Danisch}}\ and\ \bibinfo {author} {\bibfnamefont {J.}~\bibnamefont
  {Krumbiegel}},\ }\href {https://doi.org/10.21105/joss.03349} {\bibfield
  {journal} {\bibinfo  {journal} {Journal of Open Source Software}\ }\textbf
  {\bibinfo {volume} {6}},\ \bibinfo {pages} {3349} (\bibinfo {year}
  {2021})}\BibitemShut {NoStop}%
\end{thebibliography}%

\clearpage
%======================== END MATTER ========================
\renewcommand{\theequation}{S\arabic{equation}}\setcounter{equation}{0}
\renewcommand{\thefigure}{S\arabic{figure}}\setcounter{figure}{0}
\renewcommand{\thetable}{S\arabic{table}}\setcounter{table}{0}

\appendix
\section{Numerical methods}
\label{sec:numerics}

\subsection{Annealed transfer operator and Fourier--Galerkin discretization}
The annealed transfer operator is
$\mathcal{L}_\sigma=\mathcal{D}_\sigma\,\mathcal{L}$, where $\mathcal{L}$
is the Perron--Frobenius operator of the deterministic map and
$\mathcal{D}_\sigma$ is convolution with the wrapped Gaussian noise
kernel. We represent $\mathcal{L}_\sigma$ in the Fourier basis
$\{e^{2\pi i k x}\}$ of size $2K+1$, ordered
$(0,1,\ldots,K,-K,\ldots,-1)$, following the Galerkin scheme
of~\citeem{GalatoloEtAl2026}; $\mathcal{L}$ is assembled with the
\texttt{FourierAdjoint} routine of
\texttt{RigorousInvariantMeasures.jl}~\citeem{GalatoloEtAl2020,NisoliPoloni_RIM}.
In this basis $\mathcal{D}_\sigma$
is diagonal with entries $e^{-\sigma^2\pi^2 k^2/2}$, and the single
parameter $\sigma$ likewise sets the driving-noise strength in the
time-domain simulations, so that the spectral and empirical results
refer to one and the same $\sigma$.

Because the noise kernel damps the Fourier modes super-exponentially,
$\mathcal{D}_\sigma$ smooths any bounded density into a real-analytic
one, so for every $\sigma>0$ the annealed operator $\mathcal{L}_\sigma$
is compact---in fact trace-class, since $\sum_k e^{-2\sigma^2\pi^2k^2}<\infty$.
Its spectrum is therefore purely discrete, a sequence of eigenvalues
$1=\lambda_0>|\lambda_1|\ge|\lambda_2|\ge\cdots\to0$ with \emph{no
essential spectrum}; the simple leading eigenvalue $\lambda_0=1$ has
the stationary density $\rho^*$ as right eigenvector. The subleading
eigenvalues $\lambda_i$ ($i\ge1$), lying strictly inside the unit disk,
are the \emph{Ruelle--Pollicott resonances}~\citeem{Ruelle86,DyatlovZworski2015}:
the poles of the
meromorphically continued resolvent $(z-\mathcal{L}_\sigma)^{-1}$
(equivalently of the correlation spectrum), which govern the
exponential decay and oscillation of correlations
$C(n)\sim\sum_i A_i\lambda_i^{\,n}$. The noise turns what in the
noiseless limit are resonances of a meromorphic continuation into
genuine eigenvalues, resolved by the Galerkin truncation as
$K\to\infty$.

The truncation $K(\sigma)$ is set so the noise kernel is negligible at
  frequency~$K$,
  $K(\sigma)=\max\!\bigl(64,\,\mathrm{nextpow}_2\!\lceil
  \sqrt{6\ln 10}/(\pi\sigma)-\ln\sigma-\tfrac12\ln(2\pi)\rceil\bigr)$,
  which grows rapidly as $\sigma\to0$ because the kernel decays more
  slowly: for the cases of Fig.~\ref{fig:statper}
  ($\sigma=3\times10^{-4},0.007,0.009,0.014$) it gives
  $K=4096,256,256,128$ (matrix size $2K+1$), rising to $K=16384$
  (size $32769$) as $\sigma\to10^{-4}$. For the smallest noise levels
  ($\sigma\le10^{-3}$) the nominal $K(\sigma)$ is prohibitively large and
  we cap the matrix at $K=1024$. The heuristic justification is that the
  Gaussian convolution equips $\mathcal{L}_\sigma$ with a
  Doeblin--Fortet--Lasota--Yorke inequality, which controls its
spseudospectrum~\citeem{BlumenthalEtAl2026}: the dominant resonances lie
  where the pseudospectral radius is small, so they are already resolved
  at a truncation well below the nominal $K(\sigma)$, whereas only the
  dense cloud of small eigenvalues approaching the origin would require
  the full $K(\sigma)$. We stress that this is only heuristic---it
  explains why a coarse truncation captures the leading resonances but
  does not by itself certify the capped estimates.
\subsection{Computing spectral data and projectors}

Because \(\mathcal{L}_\sigma\) is non-normal, its eigenvector matrix
\(V\) is severely ill-conditioned, with
\(\kappa(V)\sim10^{13}\). Direct diagonalization followed by inversion
of \(V\) therefore yields unreliable left eigenvectors. We instead
adopt the following backward-stable procedure~\cite{GolubLoan2013}.
First, we compute the Schur factorization \(M=ZTZ^H\). We then use
\texttt{ordschur} to move the \(k\) eigenvalues of largest modulus into
the leading block,
\(T=\left(\begin{smallmatrix}T_{11}&T_{12}\\0&T_{22}\end{smallmatrix}\right)\),
and solve the Sylvester equation
\(T_{11}X-XT_{22}=T_{12}\). This yields the biorthogonal left invariant
subspace \(L=Z_1+Z_2X^H\), satisfying \(L^HZ_1=I_k\) and
\(L^HM=T_{11}L^H\). Finally, we diagonalize only the small
\(k\times k\) block \(T_{11}\). If
\(T_{11}S_1=S_1\Lambda_1\), the right and left eigenvector matrices are
\(V_R=Z_1S_1\) and \(W=LS_1^{-H}\), respectively. These provide the
discrete representations of the right and left eigenfunctions, and
satisfy \(W^HV_R=I_k\) by construction. Thus, the potentially
ill-conditioned diagonalization is confined to the small leading block,
with \(k\sim10\).

\subsection{Power spectrum}
For an observable $\phi$ the (centered) autocovariance under the
stationary measure $\rho$ is
$C(n)=\langle\phi,\mathcal{L}_\sigma^{\,n}(\bar\phi\rho)\rangle
-|\langle\phi,\rho\rangle|^2$, and the power spectrum is its
Wiener--Khinchin transform $S(\omega)=\sum_{n}C(n)e^{-i\omega n}$.
Inserting the spectral resolution
$\mathcal{L}_\sigma^{\,n}=\sum_i\lambda_i^{\,n}\,f_i\otimes\ell_i$ gives
the resonance expansion $C(n)=\sum_i A_i(\phi)\,\lambda_i^{\,n}$ with
\begin{equation}
A_i(\phi)=\underbrace{\ell_i(\bar\phi\rho)}_{\text{injection}}\;
\underbrace{\textstyle\int\bar\phi\,f_i\,dm}_{\text{visibility}} ,
\label{eq:Ai}
\end{equation}
Carrying out the Wiener--Khinchin sum term by term---each geometric
series $\sum_n\lambda_i^{\,|n|}e^{-i\omega n}$ converging because the
resonances lie strictly inside the unit disk, $|\lambda_i|<1$---gives
the power spectrum in closed form as a superposition of Lorentzians,
\begin{equation}
S(\omega)=\sum_i\frac{A_i(\phi)\,(1-\lambda_i^{2})}
{\,(1-\lambda_i e^{-i\omega})(1-\lambda_i e^{i\omega})\,},
\label{eq:SW}
\end{equation}
one term per Ruelle--Pollicott resonance---the explicit bridge between
the Wiener--Khinchin theorem and the RP spectrum: the poles of
$(1-\lambda_i e^{\pm i\omega})^{-1}$ at $\lambda_i=r_ie^{i\theta_i}$ are
the spectral lines of $S(\omega)$, each a Lorentzian at
$\omega=\theta_i$ of height $\approx 2\,\mathrm{Re}\,A_i(\phi)\,r_i/(1-r_i)$,
with $\arg\lambda_i$ the frequency, $1-r_i$ the linewidth, $A_i(\phi)$
the weight. The coefficients
follow from the Schur--Sylvester data: with $f_i,\ell_i$ the
right/left eigenvectors (columns of $V_R,W$, $W^{H}V_R=I$) and
$\bar\phi\rho$ formed alias-free on a zero-padded grid
($N=\mathrm{nextpow}_2(4K{+}2)$), one has
$A_i=[W_i^{H}(\bar\phi\rho)]\,[\phi^{H}V_{R,i}]$ (injection $\times$
visibility), $\phi$ being the observable's Fourier vector.
In practice we do not
truncate the sum: since $C(n)=\langle\phi,\mathcal{L}_\sigma^{\,n}
(\bar\phi\rho)\rangle$, we evaluate it directly by iterating the
discretized matrix on the vector $\bar\phi\rho$, which sums \emph{all}
modes and returns the exact autocovariance of the discretized operator
(Figs.~\ref{fig:statper}, using $\phi(x)=x$). All spectra are reported in
absolute units, with no rescaling between prediction and data.

\subsection{Invariant density and Lyapunov exponent.}
The invariant density $\rho$ is the $\lambda\!=\!1$ right eigenvector,
normalized to $\int\rho\,dx=1$. The annealed Lyapunov exponent is
$\Lambda=\int\log|T'|\,d\rho_\sigma
=\mathrm{Re}\,\langle\rho,\hat g\rangle$, where $\hat g$ is the
truncated Fourier representation of $\log|T'(x)|$ in the same basis and
the inner product runs over all $2K+1$ modes. Sweeping
$\sigma$, the two complex-conjugate resonance pairs form a period-3
($|\arg\lambda|\approx2\pi/3$) and a period-6 branch
($|\arg\lambda|\approx\pi/3$), tracked continuation-consistently across
their crossover; $\Lambda$ dips negative over the window in which the
period-6 resonance sharpens (Fig.~\ref{fig:lyapspec}).

Empirical power spectra are estimated using Welch's
method~\cite{Welch1967}, with Hann-windowed segments of length \(2^{13}\),
from trajectories of \(2^{20}\) iterations. After discarding an initial
transient of \(5\times10^{4}\) iterations, the spectra are averaged over
20 independent realizations and, for the deterministic reference, over
50 trajectories. The theoretical predictions and numerical results are
compared in absolute units, without any fitted multiplicative factor.
As \(\sigma\to0\), the predicted spectrum converges to the empirical
spectrum of the deterministic system, and close agreement persists
across the Lyapunov-stability transition
[Fig.~\ref{fig:statper}].
\subsection{Software.}
\sloppy
Computations were carried out in Julia~\citeem{Bezanson2017Julia} with
\texttt{RigorousInvariantMeasures.jl}~\citeem{NisoliPoloni_RIM} (operator
assembly), \texttt{IntervalArithmetic.jl}~\citeem{SandersBenet2014} and
\texttt{BallArithmetic.jl}~\citeem{FerrantiNisoli_BallArithmetic}
(verified arithmetic), \texttt{FFTW}~\citeem{JohnsonFrigo2008},
\texttt{ForwardDiff.jl}~\citeem{Revels2016ForwardDiff}, and
\texttt{Makie.jl}~\citeem{Danisch2021Makie}.

\bibliographystyleem{apsrev4-2}
\bibliographyem{lmm}

\end{document}